\documentclass[12pt]{article}
\usepackage{amsfonts,amsmath,amssymb,mathrsfs}

\title{Are All Particles Real?}
\author{
Sheldon Goldstein\footnote{Departments of Mathematics, Physics and
     Philosophy, Hill Center, Rutgers, The State University of New
     Jersey, 110 Frelinghuysen Road, Piscataway, NJ 08854-8019, USA.
     E-mail: oldstein@math.rutgers.edu},
James Taylor\footnote{Department of Mathematics, Iowa State
     University, Carver Hall, Ames, IA 50010, USA. E-mail:
     jostylr@member.ams.org},\\
Roderich Tumulka\footnote{Dipartimento di Fisica dell'Universit\`a di
     Genova and INFN sezione di Genova, Via Dodecaneso 33, 16146
     Genova, Italy. E-mail: tumulka@mathematik.uni-muenchen.de},\
and
Nino Zangh\`\i\footnote{Dipartimento di Fisica dell'Universit\`a di
     Genova and INFN sezione di Genova, Via Dodecaneso 33, 16146
     Genova, Italy. E-mail: zanghi@ge.infn.it}
}
\date{September 27, 2004}

\newcommand{\RdN}{\mathbb{R}^{3N}}
\newcommand{\RdNn}{\RdN_{\neq}}
\newcommand{\NRd}{{}^{N}\mathbb{R}^3}
\newcommand{\Rd}{\mathbb{R}^3}
\newcommand{\RRR}{\mathbb{R}}
\newcommand{\CCC}{\mathbb{C}}

\newcommand{\vn}{\boldsymbol{n}}
\newcommand{\vq}{{\boldsymbol{q}}}
\newcommand{\vQ}{{\boldsymbol{Q}}}
\newcommand{\vj}{{\boldsymbol{j}}}

\newcommand{\Laplace}{\Delta}
\newcommand{\Q}{\mathcal{Q}}

\renewcommand{\Im}{\mathrm{Im}}
\newcommand{\Indexset}{\mathscr{I}}
\newcommand{\psp}[2]{\langle #1 , #2 \rangle} 

\begin{document}\maketitle
\begin{abstract}
   In Bohmian mechanics elementary particles exist objectively, as
   point particles moving according to a law determined by a
   wavefunction. In this context, questions as to whether the
   particles of a certain species are real---questions such as, Do
   photons exist?  Electrons? Or just the quarks?---have a clear
   meaning.  We explain that, whatever the answer, there is a
   corresponding Bohm-type theory, and no experiment can ever decide
   between these theories.  Another question that has a clear meaning
   is whether particles are intrinsically distinguishable, i.e.,
   whether particle world lines have labels indicating the species. We
   discuss the intriguing possibility that the answer is no, and
   particles are points---\emph{just points}.

   \medskip

   \noindent PACS number:
   03.65.Ta (foundations of quantum mechanics)\\
   Key words: Bohmian mechanics, ontology, empirical equivalence,
   fundamental limitations of science, particle trajectories in quantum
   physics
\end{abstract}

\section{Introduction}

We address in this paper rather basic but intimidating questions about
the ontology in Bohmian mechanics and similar theories, using two
specific questions as a case study. What is intimidating about these
questions is that they cannot be answered experimentally. However, as
we shall explain, this does not mean they cannot be answered.  Most,
if not all, of what we point out in this paper has surely been known
to some experts. However, we have found no clear discussion of the
matter in the literature.

Put succinctly, Bohmian mechanics is quantum mechanics with particle
trajectories (e.g., Bohm, 1952a; Bohm, 1952b; Bell, 1966; D\"urr,
Goldstein \& Zangh\`\i, 1992; D\"urr, Goldstein \& Zangh\`\i, 1996;
Goldstein, 2001).
It describes the motion of point particles in physical space $\Rd$. In
the conventional form of Bohmian mechanics, the law of motion for the
position $\vQ_i(t)$ of the $i$-th particle of a system of $N$
particles is
\begin{equation}\label{Bohm}
   \frac{d\vQ_i}{dt} = \frac{\hbar}{m_i} \Im \frac{\psi_t^* \nabla_i
   \psi_t}{\psi_t^* \, \psi_t} \bigl( \vQ_1(t), \ldots, \vQ_N(t) \bigr)
\end{equation}
where $\psi_t: \RdN \to \CCC^\ell$ is a wavefunction obeying the
Schr\"odinger equation
\begin{equation}\label{Schroedinger}
   i\hbar \frac{\partial \psi_t}{\partial t} = - \sum_{i=1}^N
   \frac{\hbar^2}{2m_i} \Laplace_i \psi_t + V\psi_t,
\end{equation}
$m_i$ is the mass of the $i$-th particle, numerator and denominator in
\eqref{Bohm} involve scalar products in the space $\CCC^\ell$
(corresponding to spin, quark color, quark flavor and similar degrees
of freedom), $\Laplace_i$ is the Laplacian with respect to $\vq_i$
(the generic coordinates of the $i$-th particle), and finally $V$ is
the (possibly Hermitian $\ell \times \ell$-matrix-valued) potential
function on $\RdN$. The configuration $Q(t) = (\vQ_1(t), \ldots,
\vQ_N(t))$ is random with distribution $|\psi_t|^2$ at every time $t$
if $Q(0)$ is random with distribution $|\psi_0|^2$, as we shall assume
in the following (Bohm, 1952a; Bohm, 1952b; D\"urr, Goldstein \&
Zangh\`\i, 1992).
While the law of motion of Bohmian mechanics is highly non-Newtonian,
Bohmian mechanics has in common with Newtonian mechanics that there
are real particles---with actual positions---in contrast to most other
versions of quantum mechanics.

One of the two questions we want to address in this paper is whether
all elementary species (such as electron, muon, tauon, quark, photon,
gluon, etc.) actually have particles. We have to explain what we mean
by this.  When we say that certain species have \emph{no} particles,
we have in mind the following modification of the theory defined by
\eqref{Bohm} and \eqref{Schroedinger}.  Let $\Indexset \subset \{1,
\ldots, N\}$ be a nonempty index set (the set of the labels of all
``real'' particles), and stipulate that only $\# \Indexset$ particles
exist, labeled by the elements of $\Indexset$ and moving according to
\begin{equation}\label{BohmIndex}
   \frac{d\vQ_i}{dt} = \frac{\hbar}{m_i} \Im \frac{\psp{\psi}{\nabla_i
   \psi}_{\Indexset^c}}{\psp{\psi}{\psi}_{\Indexset^c}} \bigl(
   \vQ_j(t): j \in \Indexset)
\end{equation}
for $i \in \Indexset$, where $\Indexset^c = \{1, \ldots, N\} \setminus
\Indexset$, and
\begin{equation}\label{spIndex}
   \psp{\phi}{\psi}_{\Indexset^c} \, (\vq_j : j \in \Indexset) :=
   \int\limits_{\RRR^{3\#\Indexset^c}} \Bigl(\prod_{k \in \Indexset^c}
   d\vq_k \Bigr)\: \phi^*(\vq_1, \ldots, \vq_N) \, \psi(\vq_1, \ldots,
   \vq_N)
\end{equation}
is the partial scalar product between two wavefunctions $\phi$ and
$\psi$, involving the inner product in $\CCC^\ell$ and integration
over the coordinates labeled by $\Indexset^c$ (of all ``unreal
particles''), yielding a complex function of the coordinates labeled
by $\Indexset$ (of the ``real particles''). Equations
\eqref{BohmIndex} and \eqref{spIndex} are completed by the unchanged
Schr\"odinger equation \eqref{Schroedinger}.

This is a theory of $\#\Indexset$ particles, even though
\eqref{Schroedinger} looks as if it is about $N$ particles. We could
take $\Indexset$ to contain all electrons but not the quarks, or
instead all quarks but not the electrons. As we will point out in
Section~2, for both of these and many other choices, no conceivable
experiment can detect a difference from conventional Bohmian mechanics
as defined by \eqref{Bohm} (or from one another), provided that the
universe is in quantum equilibrium\footnote{In order to appreciate the
strength of this indistinguishability, it is important to recognize
that quantum equilibrium is absolute, not in the sense that the
universe has to be in quantum equilibrium---it doesn't, although the
correctness of the quantum mechanical predictions strongly suggests
that it is---but in the sense that a universe in quantum equilibrium,
like one in thermodynamic equilibrium, is pretty much stuck there and
nothing we can do, nothing we can control, can get us out of
it. Nonetheless, quantum non-equilibrium versions of these theories
would be empirically distinguishable, from each other and from
orthodox quantum theory; this sort of possibility has been explored by
Valentini (2002).
} (see D\"urr, Goldstein \& Zangh\`\i, 1992)
so that the configuration of the (real) particles has probability
distribution
\begin{equation}\label{rhoIndex}
    \rho = \psp{\psi}{\psi}_{\Indexset^c}.
\end{equation}
We will also indicate in Section~2 why \eqref{Bohm} may be
more plausible than \eqref{BohmIndex}. Nevertheless, we believe that the
impossibility, in quantum equilibrium, of conclusively rejecting
\eqref{BohmIndex} in favor of \eqref{Bohm}, or \eqref{Bohm} in favor of
\eqref{BohmIndex}, is but another instance of one of the \emph{fundamental
limitations of science}: that there may be distinct theories that are
empirically indistinguishable.

The second question we consider in this paper is whether the
configuration space should be, instead of $\RdN$,
\begin{equation}\label{NRddef}
   \NRd := \bigl\{ S \subseteq \Rd : \# S = N \bigr\},
\end{equation}
the space of all $N$-element subsets of $\Rd$. $\NRd$ can also be
identified with $\RdNn$ modulo permutations, where
\begin{equation}\label{RdNndef}
   \RdNn := \bigl\{ (\vq_1, \ldots, \vq_N) \in \RdN: \vq_i \neq \vq_j
   \: \forall i \neq j \bigr\}
\end{equation}
is $\RdN$ minus the coincidence configurations. The choice of $\NRd$
as configuration space corresponds to the notion that a configuration
of $N$ particles is a set of $N$ points in physical space, with the
points labeled in no way, neither by numbers $1, \ldots, N$, nor in
the sense that there could be intrinsically different kinds of points
in the world, such as electron points as distinct from muon points or
quark points. (The configuration space $\NRd$ has nontrivial
topology. See D\"urr, Goldstein, Taylor, Tumulka \& Zangh\`\i\ (2005)
for an investigation of the implications of this fact for Bohmian
mechanics. This space has been suggested as the configuration space of
$N$ identical particles by Laidlaw \& DeWitt (1971), Leinaas \&
Myrheim (1977), Nelson (1985), Brown, Sj\"oqvist \& Bacciagaluppi
(1999), and D\"urr, Goldstein, Taylor, Tumulka \& Zangh\`\i\ (2005).)

To be specific, we have in mind again a concrete Bohm-type theory,
combining the usual Schr\"odinger equation \eqref{Schroedinger} with a
modification of \eqref{Bohm}. Equation \eqref{Bohm} itself does not
define a dynamics for an unordered configuration, from $\NRd$, because
different ways of numbering the $N$ given points would (generically)
lead to different trajectories of the particles. To obtain a
well-defined dynamics on $\NRd$, we note that the right hand side of
\eqref{Bohm} can be written as
\begin{equation}
   \frac{\vj_i}{\rho} \bigl( \vQ_1(t), \ldots, \vQ_N(t) \bigr)
\end{equation}
with
\begin{equation}\label{jdef}
   \vj_i = \frac{\hbar}{m_i} \Im \, \psi^* \nabla_i \psi
\end{equation}
and
\begin{equation}\label{rhodef}
   \rho = \psi^* \, \psi.
\end{equation}
Now symmetrizing $j=(\vj_1, \ldots, \vj_N)$ and $\rho$ leads to a new
equation of motion
\begin{equation}\label{Bohmsym}
   \frac{d\vQ_i}{dt} = \frac{\sum\limits_{\sigma \in S_N}
   \vj_{\sigma(i)} \circ \sigma} {\sum\limits_{\sigma \in S_N} \rho
   \circ \sigma} \bigl( \vQ_1(t), \ldots, \vQ_N(t) \bigr)
\end{equation}
where $S_N$ is the set of permutations of $\{1, \ldots, N\}$ and
\[
\sigma(\vQ_1, \ldots, \vQ_N) := (\vQ_{\sigma^{-1}(1)}, \ldots,
\vQ_{\sigma^{-1}(N)})
\]
so that $\vQ_j$ is moved to the place $\sigma(j)$. Observe that
renumbering the particles now does not change the
trajectories.\footnote{Let us connect the transition from \eqref{Bohm}
to \eqref{Bohmsym} with the ``problem of recognition'' raised by
Brown, Elby, and Weingard (1996, p.~314) 
in a paper that is largely concerned with whether particle masses
should be attributed to particle locations, or to the wave function,
or to both---a question that does not concern us.  Be that as it may,
their argument concerning the problem of recognition, as we understand
it, amounts to the observation that the labeled particles in the
conventional Bohmian mechanics of distinguishable particles indeed
{\em must be} labeled, in the sense of being intrinsically
metaphysically distinct: otherwise which particle corresponds to which
argument of the wave function and, consequently, to which velocity
would be ambiguous.  We note that it is exactly this ambiguity that
renders \eqref{Bohm} inadequate for unordered or unlabeled
configurations, and that it is precisely this ambiguitity that is
removed by the symmetrized dynamics \eqref{Bohmsym}: with
\eqref{Bohmsym} the positions of the particles alone suffice for
deciding which one has which velocity.}

The difference between \eqref{Bohm} and \eqref{Bohmsym}, which lead to
different trajectories and thus define inequivalent theories, makes
clear that in Bohmian mechanics the question whether particles are
\emph{just points}, represented by an element of $\NRd$, or
\emph{labeled points}, represented by an element of $\RdN$, has a
genuine physical meaning. Again, the question can never be decided
experimentally, as we will explain in Section~3. There we will also
say why this time the conventional formula \eqref{Bohm} may be less
plausible than its modification.

\section{Do Photons Exist? Electrons? Quarks?}

A relevant mathematical fact about Bohmian mechanics with a reduced
number $\# \Indexset$ of particles, as defined by \eqref{BohmIndex},
is the equivariance of the probability distribution \eqref{rhoIndex}:
if the configuration $Q(t) = (\vQ_j(t) : j \in \Indexset)$ is random
with distribution $\rho_t = \psp {\psi_t} {\psi_t}_{\Indexset^c}$ at
some time $t$, then this is so for all times $t$. This follows from
the continuity equation for $\psi^* \psi$ implicit in the
Schr\"odinger equation \eqref{Schroedinger} by integrating over the
coordinates $\vq_k$ for $k \in \Indexset^c$.

The distribution \eqref{rhoIndex} is the basis of the empirical
equivalence with conventional Bohmian mechanics as defined by
\eqref{Bohm}: Suppose that the result of an experiment is recorded in
the positions $\vQ_j$ of the ``real'' particles, $j \in \Indexset$.
Since their distribution is the same as the marginal distribution of
the $\vQ_j$ in conventional Bohmian mechanics,
\begin{equation}\label{marginals}
   \psp{\psi_t}{\psi_t}_{\Indexset^c} (\vq_j: j \in \Indexset) =
   \int\limits_{\RRR^{3 \# \Indexset^c}} \Bigl(\prod_{k \in
   \Indexset^c} d\vq_k \Bigr)\: \bigl|\psi_t(\vq_1, \ldots,
   \vq_N)\bigr|^2,
\end{equation}
there is no way to tell from the values of $\vQ_j$, $j \in \Indexset$,
at one particular time whether they were generated using the reduced
dynamics \eqref{BohmIndex} or the conventional one \eqref{Bohm}.  The
same is true of any sequence of experiments, wherein the choice of the
second experiment may even depend on the outcome of the first (D\"urr,
Goldstein \& Zangh\`\i, 1992). Thus, no conceivable experiment can
decide between the two theories, provided that the outcomes of any
conceivable experiment always get recorded in the $\vQ_j$, $j \in
\Indexset$.\footnote{And provided that the $\vQ_j$, $j \in \Indexset$,
always contain, in addition to the outcome, sufficient information to
judge whether the experiment was properly conducted. All of this
information would be available if the $\vQ_j$, $j \in \Indexset$,
suffice to define the ``macroscopic configuration.''

   Aside from this, there is a subtlety here that need not bother us
   for our considerations but should be mentioned nevertheless. It
   does not quite follow from equivariance that the configurations
   after the experiment will have the same distribution. This is
   illustrated by the following example that we owe to Tim~Maudlin
   (private communication): consider Nelson's (1985) stochastic
   mechanics (more precisely, consider the formulation of stochastic
   mechanics due to Davidson (1979) for which the diffusion
   coefficient is a free parameter) with \emph{extremely} large
   diffusion coefficient. Since the wave function of the universe must
   presumably be thought of as consisting of several packets that are
   very far apart in configuration space $\RRR^{3N}$, such as packets
   that correspond to unrealized outcomes of quantum measurements, the
   configuration $Q(t)$ will very probably also visit---in every
   second---those distant regions supporting the other packets, in
   some of which the dinosaurs have never become extinct.  In this
   case the conditional distribution for the configuration after the
   experiment, conditional on that we did begin to perform the
   experiment, will be drastically different in stochastic mechanics
   from what it is in Bohmian mechanics. It is still the case that
   discrimination will not be possible---we probably will not even
   remember that we did that experiment in stochastic mechanics---but
   not exactly because of the simple reason that comes to mind at
   first.}

An example of such a set $\Indexset$ would be the set of all quarks.
Pointer positions, ink marks on paper, and even the memory contents in
a brain are reflected by the configuration of quarks. The same is true
of electrons: if we know of an atom where its electrons are, we
roughly know where its nucleus would have been if it existed. Even the
positions of all photons alone would presumably suffice for fixing the
macroscopic configuration, as every electron is believed to be
surrounded by a cloud of photons, so that the photon configuration
would roughly define the electron configuration (for quantum states
relevant to ordinary macroscopic bodies) well enough to determine the
values of macroscopic variables.\footnote{Here, too, there is a
   sublety into which we don't wish to delve in this paper. The point
   is that there is a sense in which, arguably, experimental
   discrimination is trivial in the only-photons version, because the
   predictions come out drastically wrong. It's not simply that
   pointers don't end up pointing with the right distribution; it could
   be argued that pointers don't end up pointing at all because there
   are no pointers---after all, pointers are not usually considered to
   be made of photons. Thus the only-photons theory would then seem to
   make no decent predictions at all. Of course a proponent of the
   only-photons theory wouldn't be inclined to accept that conclusion:
   he would, perhaps, appeal to the information encoded in the pattern
   of photon positions as the foundation of the predictions of his
   theory, using a map from photon configurations to (in fact,
   non-actual) electron configurations to results of experiments. For
   such a person, regardless of whether he is right or wrong in
   behaving as he does, experimental discrimination between his theory
   and a more normal theory would be impossible.}

We encounter here one of the fundamental limitations of science.
There is no way of experimentally testing various possibilities for
$\Indexset$ against each other. There is no way of \emph{conclusively
establishing} what the true $\Indexset$ is. This situation is probably
new for most physicists;\footnote{The situation is in fact not new to
researchers in Bohmian mechanics, as Bohmian mechanics is known to be
empirically equivalent to other dynamics on $\RRR^{3N}$ that make the
$|\psi|^2$ distribution equivariant, such as Nelson's (1985)
stochastic mechanics and the velocity laws studied by Deotto and
Ghirardi (1998).} most of us are used to contemplating rival theories
that make similar predictions, but not rival theories that make
exactly and perfectly the same predictions for all experiments. But
the conclusion is simple: we have no alternative to accepting that we
cannot finally know what $\Indexset$ is, and we should simply admit
this.

Nevertheless, this does not mean that all possibilities are equally
plausible. On the contrary, many of the possibilities are ridiculous.
For example, it is ridiculous to assume that everything presently
outside the solar system is not real, though we (presently living
humans) would not be able to find out (since outcomes of our
experiments would be recorded in the particles inside the solar
system). As another example, let $\Indexset$ be a typical set with
roughly $N/2$ elements; what makes this ridiculous is the complete
arbitrariness of what is real and what is not. If we relax the
assumption that $\Indexset$ is a \emph{fixed} set (how this can be
done we describe in Appendix~A), we find even more drastic examples:
that everything outside the United States is not real (people in the
US would not be able to find out), or that women are not real (men
would not be able to find out), or---as a solipsistic kind of Bohmian
theory---that only your brain, dear reader, is real (you would not be
able to find out).  Thus, some possibilities would be rejected by
everyone, on basically the same grounds that one would reject
solipsism: though it is logically possible and experimentally
irrefutable, it is implausible---or whichever word you prefer.

It seems quite ridiculous as well, though perhaps not as obviously, to
assume that quarks are not real. A Bohmian world in which quarks are
not real would be an eccentric world, and a God who created such a
world would certainly count as malicious. The only world that seems to
us not eccentric at all is the one in which all particles are real. We
mention some reasons: The Schr\"odinger equation \eqref{Schroedinger}
suggests that there are $N$ particles. Why should there be coordinates
in the wave function, varying in physical space, if there is no
corresponding particle? On top of that, that all particles are real
seems to be everybody's intuition, so much so that literally all
physics texts talk about photons, electrons, and quarks, even those
maintaining that microscopic reality does not exist. Another thing to
keep in mind is that you may have found it difficult back in Section~1
to understand the strange thought that some particles are ``real'' and
some are ``not real''; in any case, we found it difficult to explain
and even to merely express. This may indicate that \eqref{BohmIndex}
is not a very natural idea.

We finally remark that one could perhaps imagine that there may be
compelling reasons (of a mathematical or physical kind) precluding
certain species from having particles. These would of course be
reasons for preferring \eqref{BohmIndex} to \eqref{Bohm}; presently,
however, we do not know of any such reasons.

\section{Are Particles Just Points?}

We now turn to a discussion of the symmetrized law of motion
\eqref{Bohmsym}. The empirical equivalence between symmetrized Bohmian
mechanics, taking place in $\NRd$, and conventional Bohmian mechanics
as defined by \eqref{Bohm} is again based on equivariance.  As is
easily checked, \eqref{Bohmsym} implies equivariance of the
distribution
\begin{equation}\label{rhosym}
   \rho_\mathrm{sym} = \sum_{\sigma \in S_N} \rho \circ \sigma\,,
\end{equation}
a symmetric distribution on $\RRR^{3N}$ which thus defines a
distribution on $\NRd$, which again equals a marginal of the
distribution \eqref{rhodef} of the conventional configuration $(\vQ_1,
\ldots, \vQ_N) \in \RRR^{3N}$, namely the distribution of the set
$\{\vQ_1, \ldots, \vQ_N\}$ disregarding the labels. Now empirical
equivalence follows, by a reasoning similar to the one following
\eqref{marginals}, from the fact that outcomes of all conceivable
experiments will be recorded in the unordered configuration $\{\vQ_1,
\ldots, \vQ_N\}$. To illustrate this fact, we may imagine the outcome
as given by the orientation of a pointer on a scale; as the pointer
consists of a huge number of electrons and quarks, for reading off the
orientation of the pointer we need not be explicitly told which points
are the electrons and which are the quarks.  Moreover, it is important
to bear in mind that our assessment of which particles are quarks and
which are electrons presumably could not be based on any sort of
direct access to the particle's intrinsic nature, but rather must be
based on information about the particle's behavior, reflected in the
overall configuration of the particles.

We have two possibilities: particles belonging to different species
may be metaphysically different, i.e., electron points may be
different from quark points or photon points, or, alternatively, they
may all be just points.  As in the situation of the previous section,
the impossibility of deciding experimentally between these
possibilities is a fundamental limitation of science. A choice can
only be based on theoretical considerations, and this time both
possibilities seem plausible enough to be acceptable.

We would like to point out, though, that the possibility that all
particles are just points, first implicitly suggested in Bell's (1986)
seminal paper, is quite attractive, more attractive than it may appear
at first. And not merely because of metaphysical simplicity. As we
have already indicated, if, say, electrons and muons were different
kinds of points then this difference in the nature of these points
would be in no way directly accessible to us. Our decision as to
whether a given particle is an electron or a muon would be based on
its behavior under the conditions (such as external fields) we impose,
i.e., based on its trajectory.

We add another thought. Sometimes progress in theoretical physics
forces us to regard what was previously considered two species as two
quantum states of the same species. For instance, proton and neutron
appeared as two species, but in fact are two states of a three-quark
system. The most radical development possible in this direction,
considered by Goldstein, Taylor, Tumulka \& Zangh\`\i\ (2004), would
be that all of what we presently consider different species are just
different states of the same species---which would of course force us
to adopt the symmetrized dynamics \eqref{Bohmsym}, and would suggest
against \eqref{BohmIndex}.

\section{Conclusions}

We have made the observation that the following two questions have a
clear physical meaning in Bohmian mechanics, however obscure they may
be in other versions of quantum mechanics: Do all particle species
have particles? And, do the particles have labeled or unlabeled world
lines?  These questions are about various conceivable equations of
motion that we have explicitly specified. We have underlined that the
various theories are empirically equivalent (provided the set of real
particles is not too small), so that any answer to these questions
would have to be grounded in purely theoretical, possibly
philosophical, considerations.  We have argued that the most
convincing answer to the first question is that all species have
particles, unless this is precluded by compelling mathematical or
physical reasons yet to be discovered. Concerning the second question,
we have suggested that the most attractive possibility is that world
lines are unlabeled: what are normally regarded as distinguishable
particles are better regarded as intrinsically indistinguishable.

Bell (1986) expressed concern 
about the empirical equivalence of various choices of what is real, in
the context of his lattice model analogous to Bohmian mechanics. The
situation may be much better, however, than Bell thought: concerning
our question as to which particles are real, the possibilities differ
greatly in plausibility, and one of them seems clearly more natural
than all others---so there is not so much to be concerned about.

\section*{Appendix A}

Here is a brief outline of how the dynamics \eqref{BohmIndex} can be
modified in order to deal with a variable set $\Indexset$. One example
to have in mind is that, so to speak, $\Indexset(t)$ contains all
particles in a particular region $R \subseteq \RRR^3$ of physical
space. More generally, we can let $\Indexset$ depend on time and even
on the configuration.

To this end, we start with a partition $\bigl\{S_\Indexset : \Indexset
\subseteq \{1, \ldots, N\} \bigr\}$ of configuration space-time
$\RRR^{3N} \times \RRR$ indexed by all sets of particle labels.
Equivalently, we can start with a configuration-dependent set
$\Indexset(q,t) \subseteq \{1, \ldots, N\}$ indicating which particles
are real, specified by a function from configuration space-time
$\RRR^{3N} \times \RRR$ to the power set of $\{1, \ldots, N\}$; the
partition is then formed by the level sets of this function.  In the
example case in which only the particles in the region $R$ are real,
we would set $S_\Indexset = \{(\vq_1, \ldots, \vq_N,t) : \vq_j \in R,
\vq_k \notin R \text{ for } j\in \Indexset, k \in \Indexset^c \}$.
Since the number of real particles can change, the dynamics of the
configuration takes place in the disjoint union\footnote{In what
   follows, ${(\RRR^3)}^{\Indexset} = \{(\vq_j: j \in \Indexset)\}$,
   with each $\vq_j\in\RRR^3$, is the set of functions from $\Indexset$
   to $\RRR^3$. ${(\RRR^3)}^{\Indexset}$ is isomorphic to $\RRR^{3 \#
     \Indexset}$, but
   ${(\RRR^3)}^{\Indexset}\neq{(\RRR^3)}^{\Indexset'}$ even for $\#
   \Indexset = \# \Indexset'$ (unless of course $\Indexset =
   \Indexset'$).}
\[
\Q = \bigcup_{\Indexset \subseteq \{1, \ldots, N\}}
{(\RRR^3)}^{\Indexset}.
\]
Let $\pi_\Indexset: S_\Indexset \to {(\RRR^3)}^{\Indexset} \times
\RRR$ be the projection $\pi_\Indexset(\vq_1, \ldots, \vq_N,t) =
(\vq_j: j \in \Indexset,t)$, and $\pi: \RRR^{3N} \times \RRR \to \Q
\times \RRR$ the combination of the $\pi_\Indexset$, i.e., $\pi(q,t) =
\pi_\Indexset(q,t)$ when $(q,t) \in S_\Indexset$.

The dynamics for $Q(t)$ in $\Q$ we now define is the Markovization of
the stochastic process $\pi(\tilde{Q}(t),t)$, where $\tilde{Q}(t)$
denotes the conventional Bohmian motion in $\RRR^{3N}$ defined by
\eqref{Bohm}.  $Q(t)$ will consist of deterministic trajectories
interrupted by stochastic jumps that change the number (or labels) of
the real particles. See D\"urr, Goldstein, Tumulka \& Zangh\`\i\
(2004) for the general theory of particle creation in Bohmian
mechanics by means of stochastic jumps.  The deterministic motion is
defined by
\begin{equation}\label{Bohmpartition}
   \frac{d\vQ_i}{dt} = \frac{\vj_i(Q(t),t)} {\rho(Q(t),t)},
\end{equation}
where, for $q\in{(\RRR^3)}^{\Indexset}$,
\begin{equation}
   \vj_i(q,t) = \int\limits_{\pi_\Indexset^{-1}(q,t)} \Bigl(\prod_{k
   \in \Indexset^c} d\vq_k \Bigr)\: \tfrac{\hbar}{m_i} \, \Im \,
   \psi^*_t \nabla_i \psi_t
\end{equation}
is the probability current and
\begin{equation}\label{rhopartition}
   \rho(q,t) = \int\limits_{\pi_\Indexset^{-1}(q,t)} \Bigl(\prod_{k \in
   \Indexset^c} d\vq_k \Bigr)\: \psi^*_t \psi_t
\end{equation}
is the probability density on $\Q$.  The rate of jumping from $Q \in
\Q$ to a configuration $q' \in \Q$ (in which the particles from
$\Indexset'$ are real) is
\begin{equation}\label{ratepartition}
   \sigma_t(dq',Q) = \frac{J^+(dq',dQ,dt)}{\rho(Q,t) \, dQ \, dt},
\end{equation}
where $x^+ = \max(x,0)$, and $J(dq',dQ,dt)$ is the probability flow,
in $\RRR^{3N} \times \RRR$, in the time element $dt$ from
$S_\Indexset$ to $S_{\Indexset'}$ across the boundary $\partial
S_\Indexset$ at volume elements $dq'$ and $dQ$,
\begin{equation}
   J(dq',dQ,dt) = \int\limits_{D(dq',dQ,dt)} \hspace{-6mm} dA \: \Bigl(
   n_0 \psi^*_t \psi_t + \sum_{i =1}^N \tfrac{\hbar}{m_i} \, \Im \,
   \psi^*_t \vn_i \cdot \nabla_i \psi_t \Bigr)
\end{equation}
with integration domain $D(dq',dQ,dt) = \pi_\Indexset^{-1}(dQ \times
dt) \cap \pi_{\Indexset'}^{-1}(dq'\times dt) \cap \partial S_\Indexset
\cap \partial S_{\Indexset'}$, $dA$ the surface element on the
boundary and $(n_0,\vn_1, \ldots, \vn_N)$ the unit normal vector on
the boundary pointing from $S_\Indexset$ to $S_{\Indexset'}$.

By standard arguments (D\"urr, Goldstein, Tumulka \& Zangh\`\i, 2004),
one shows that the distribution \eqref{rhopartition} is equivariant.


\section*{References}
\setlength{\parindent}{0mm}

   Bell, J. S. (1966). On the problem of hidden
   variables in quantum mechanics. \textit{Reviews of Modern Physics},
   \textbf{38}, 447--452.  Appeared also in Bell (1987), p.~1.

   Bell, J. S. (1986). Beables for quantum field
   theory. \textit{Physics Reports}, \textbf{137}, 49--54. Appeared
   also in Bell (1987), p.~173.

   Bell, J. S. (1987).  \textit{Speakable and
   unspeakable in quantum mechanics}.  Cambridge: Cambridge University
   Press.

   Bohm, D. (1952a). A Suggested Interpretation of the
   Quantum Theory in Terms of ``Hidden'' Variables,
   I. \textit{Physical Review}, \textbf{85}, 166--179.

   Bohm, D. (1952b). A Suggested Interpretation of the
   Quantum Theory in Terms of ``Hidden'' Variables,
   II. \textit{Physical Review}, \textbf{85}, 180--193.

  Brown, H., Elby, A., \& Weingard, R. (1996). Cause and
  effect in the pilot-wave interpretation of quantum mechanics.  In
  J. T. Cushing, A. Fine \& S. Goldstein (Eds.), \textit{Bohmian
  Mechanics and Quantum Theory: An Appraisal}
  (pp. 309--319). Dordrecht: Kluwer Academic.

  Brown, H., Sj\"oqvist, E., \& Bacciagaluppi,
  G. (1999).  Remarks on identical particles in de Broglie--Bohm
  theory.  \textit{Physics Letters A}, \textbf{251}, 229--235.

   Davidson, M. (1979). A generalization of the
   F\'enyes-Nelson stochastic model of quantum mechanics.
   \textit{Letters in Mathematical Physics}, \textbf{3}, 271--277.

  Deotto, E., \& Ghirardi, G.C. (1998). Bohmian
  Mechanics Revisited.  \textit{Foundations of Physics}, \textbf{28},
  1--30.

  D\"urr, D., Goldstein, S., Taylor, J., Tumulka, R.,
  \& Zangh{\`\i}, N. (2005). Bosons, Fermions, and the Topology of
  Configuration Space in Bohmian Mechanics. (in preparation)

   D\"urr, D., Goldstein, S., Tumulka, R., \&
   Zangh{\`\i}, N. (2004). Quantum Hamiltonians and Stochastic
   Jumps. To appear in \textit{Communications in Mathematical
   Physics}. Preprint online arXiv.org/quant-ph/0303056

   D\"urr, D., Goldstein, S., \& Zangh\`\i, N. (1992).
   Quantum Equilibrium and the Origin of Absolute Uncertainty.
   \textit{Journal of Statistical Physics}, \textbf{67}, 843--907.

  D{\"u}rr, D., Goldstein, S., \& Zangh{\`{\i}}, N. (1996).
  Bohmian Mechanics as the Foundation of Quantum Mechanics.  In
  J. T. Cushing, A. Fine \& S. Goldstein (Eds.), \textit{Bohmian
  Mechanics and Quantum Theory: An Appraisal}
  (pp. 21--44). Dordrecht: Kluwer Academic.

   Goldstein, S. (2001). Bohmian Mechanics. In
   E.~N.~Zalta (Ed.), \textit{Stanford Encyclopedia of Philosophy}.
   Published online http://plato.stanford.edu/entries/qm-bohm

   Goldstein, S., Taylor, J., Tumulka, R., \&
   Zangh{\`\i}, N. (2004). Are All Particles Identical? Preprint
   online arXiv.org/quant-ph/0405039.

  Laidlaw, M. G., \& DeWitt, C. M. (1971). Feynman
  functional integrals for systems of indistinguishable particles.
  \textit{Physical Review D}, \textbf{3}, 1375--1378.

  Leinaas, J., \& Myrheim, J. (1977). On the theory of
  identical particles. \textit{Il Nuovo Cimento}, \textbf{37B}, 1--23.

   Nelson, E. (1985). \textit{Quantum Fluctuations}.
   Princeton: Princeton University Press.

  Valentini, A. (2002). Subquantum Information and
  Computation. \textit{Pramana - Journal of Physics}, \textbf{59},
  269--277.

\end{document}